\documentclass[aps,twocolumn,showpacs,amsmath,amssymb,superscriptaddress]{revtex4}
\usepackage{graphicx}
\usepackage{dcolumn}
\usepackage{bm}
\usepackage{hyperref}

\begin{document}

\title{High-spin binary black hole mergers}

\author{Pedro Marronetti}
\affiliation{Department of Physics,
	Florida Atlantic University, 
	Boca Raton, FL 33431, USA}
\author{Wolfgang Tichy}
\affiliation{Department of Physics,
	Florida Atlantic University, 
	Boca Raton, FL 33431, USA}
\author{Bernd Br\"ugmann}
\affiliation{Theoretical Physics Institute,
	University of Jena,
	07743 Jena, Germany}
\author{Jose Gonz\'alez}
\affiliation{Instituto de F\'isica y Matem\'aticas,
	Universidad Michoacana de San Nicol\'as de Hidalgo,
	Morelia, Mexico}
\affiliation{Theoretical Physics Institute,
	University of Jena,
	07743 Jena, Germany}
\author{Ulrich Sperhake}
\affiliation{Theoretical Physics Institute,
	University of Jena,
	07743 Jena, Germany}
	

%
\newcommand\ba{\begin{eqnarray}}
\newcommand\ea{\end{eqnarray}}
\newcommand\et{{\it et al.}}
\newcommand\remove{{{\bf{THIS FIG. OR EQS. COULD BE REMOVED}}}}
%

\begin{abstract}

We study identical mass black hole binaries with spins perpendicular to the
binary's orbital plane. These binaries have individual spins ranging from
$s/m^2=-0.90$ to $0.90$, ($s_1 = s_2$ in all cases) which is near the limit 
possible with standard Bowen-York puncture initial data. 
The extreme cases correspond to the largest initial spin simulations to date. 
Our results expand the parameter space
covered by Rezzolla {\it et al.} and, when combining both data sets,
we obtain estimations for the 
minimum and maximum values for the intrinsic angular momenta of the remnant of 
binary black hole mergers of $J/M^2=0.341 \pm 0.004$ and $0.951 \pm 0.004$ 
respectively. Note, however, that these values are reached through 
extrapolation to the singular cases $|s_1| = |s_2| = 1$ and thus
remain as {\it estimates} until full-fledged numerical simulations
provide confirmation.

\end{abstract}

\maketitle

\section{Introduction}

The existence of black holes, originally introduced as a family of solutions
to the vacuum Einstein field equations, was a matter of speculation for the
best part of the 20th century. In the past decade, however, astronomical
observations placed them as the most promising models for objects detected
in X-ray binaries (with sizes of a few to tens of solar masses) and for the
supermassive entities at the center of some galaxies (with millions to
billions of solar masses) \cite{Narayan:2005ie}. Any black hole can be fully
specified by its mass, angular momentum and charge. Since electrically
charged black holes are quickly neutralized by free charges found in their
vicinity (i.e, from accretion disks, interstellar plasma, etc.), only their
mass and angular momentum are of astrophysical relevance. In a dynamical
environment, several factors can determine the black hole's angular
momentum: the characteristics of its progenitor (in case of stellar collapse
formation), the merger with other black holes and neutron stars of
comparable mass, the merger with smaller objects, and accretion of matter
from a surrounding disk. (For a review on how these situations can alter the
black hole rate of rotation and the bounds for the maximum spin attainable
see, for instance, \cite{Gammie:2003qi} and references therein.) Highly spinning 
black hole binaries are of particular interest given their  astrophysical 
relevance. X-ray spectroscopic studies of accretion disks around the supermassive 
galactic black holes may provide evidence of spin parameters larger than $0.9$ 
\cite{Reynolds:2004qk} (however, see for instance \cite{Nandra:2006mf} on
how this may only be a tentative estimate).

Recent progress in numerical relativistic simulations of binary black holes
(BBH) \cite{Pretorius:2005gq, Campanelli:2005dd, Baker:2005vv} makes now
possible long and stable evolutions that were impractical a few years ago.
In this paper, we studied identical black holes with spins
perpendicular to the orbital plane. In general, black hole spins would have
arbitrary directions. However, it has been recently suggested
\cite{Bogdanovic:2007hp} that supermassive galactic BBH may favor
configurations with spin alignments like the ones studied here, due to the
dynamical interaction of the holes with their galactic environments. Several
groups have recently studied BBH with spins perpendicular to the orbital
plane \cite{Campanelli:2006fy, Herrmann:2007ac, Koppitz:2007ev, 
Pollney:2007ss, Rezzolla:2007xa}. Here, we
extend those studies, with binaries with initial individual spins ranging
from $s/m^2=-0.90$ to $0.90$ (these extreme values are the largest modeled
to date) with $m_1 = m_2$ and $s_1 = s_2$. The binaries in this sequence have 
not been considered before in such detail. Note also that, due to the symmetry 
of these systems, no gravitational recoil is present in the merger remnant.

We perform a least-square data fit of our data following the quadratic formula
used by Campanelli \et ~\cite{Campanelli:2006fy} and Rezzolla \et 
~\cite{Rezzolla:2007xa} and also of both Rezzolla \et ~and 
our data sets combined. The last fit predicts minimum and maximum values
for the spin parameter of the black hole remnant of 
$0.341 \pm 0.004$ and $0.951 \pm 0.004$ respectively.

We find that current numerical techniques for the evolution of black holes
with spins $s/m^2 > 0.75$ are limited in that they
produce an artificial loss of angular momentum
that increases with the magnitude of the spin. While this effect can be
diminished by increasing the grid resolution, even relatively large
resolutions such as $M/85$ present loss rates larger than $1\%$ per $100M$
for the case $s/m^2=0.90$. This effect is clearly seen in the long term
evolution of BBH that results in a highly spinning black hole and also on
single black hole simulations.

In Sec. \ref{techniques}, we present the numerical details and tests of our
simulations. Sections \ref{results} and \ref{conclusions} present our
results and conclusions.

\section{Numerical techniques and tests}
\label{techniques}

\begin{table}
{\small
\begin{tabular}{c|c|c|c|c|c|c}
$s/m^2$	& $m/M$  & $m_b/M$ & $D/M$  & $P/M$   & $M_i/M$	& $J_i/M^2$\\
\hline
-0.90	& 0.5000 & 0.1767  & 6.6965 & 0.14162 & 0.989 & 0.498 \\
-0.75	& 0.5000 & 0.3307  & 6.6372 & 0.14041 & 0.988 & 0.557 \\
-0.50	& 0.5000 & 0.4246  & 6.5366 & 0.13838 & 0.987 & 0.655 \\
-0.25	& 0.5000 & 0.4656  & 6.4337 & 0.13631 & 0.986 & 0.752 \\
 0.00	& 0.5000 & 0.4777  & 6.3286 & 0.13419 & 0.986 & 0.849 \\
 0.25	& 0.5000 & 0.4654  & 6.2212 & 0.13204 & 0.985 & 0.940 \\
 0.50	& 0.5000 & 0.4240  & 6.1117 & 0.12983 & 0.985 & 1.043 \\
 0.62	& 0.5000 & 0.3888  & 6.0583 & 0.12875 & 0.985 & 1.090 \\
 0.75	& 0.5000 & 0.3299  & 6.0000 & 0.12756 & 0.985 & 1.140 \\
 0.82	& 0.5000 & 0.2810  & 5.9684 & 0.12691 & 0.985 & 1.167 \\
 0.90	& 0.5000 & 0.1764  & 5.9320 & 0.12616 & 0.985 & 1.198 \\
\end{tabular}
}
\caption{\label{punc_par_tab}
Initial data parameters. Here $m_b$ is the bare mass parameter of each
puncture and $M = 2 m$ is the sum of the ADM masses $m$ measured at each
puncture.  The holes have coordinate separation $D$, with puncture locations 
$(0, \pm D/2 ,0)$, linear momenta $(\mp P, 0, 0)$, and spins $(0, 0, s)$. We 
also list the initial values of the ADM mass $M_i$ and the angular momentum 
$J_i$. The 2PN angular velocity 
is set to $\Omega M =0.05550$ in each case.
}
\end{table}

All the binary systems studied here consist of identical black holes with spins 
aligned with the orbital angular momentum. In our coordinates, the orbit 
develops in the $xy$ plane and the only non-vanishing component of the spins 
is in the $\hat{z}$ direction. We simulated binaries with spin parameters 
ranging from $s/m^2=-0.90$ to $0.90$.

As starting points of our simulations, we use standard puncture 
initial data~\cite{Brandt97b} with the momentum and spin parameters
in the extrinsic curvature given by second-order post-Newtonian
(2PN) estimates~\cite{Kidder:1995zr}.
It is sufficient to use 2PN estimates because standard puncture data
are inconsistent with PN theory beyond 
$(v/c)^3$~\cite{Tichy02,Yunes:2006iw,Yunes:2005nn,Kelly:2007uc}.
These parameters along with the initial 
ADM mass $M_i$ and angular momentum $J_i$
are shown in Table~\ref{punc_par_tab}.
The coordinate distance $D$, the linear momenta $P$ and spin parameters 
$s$ are directly used in the Bowen-York extrinsic curvature, while the bare 
mass parameter is obtained from the condition that
the ADM masses measured at each puncture should be $m=M/2$.
As in~\cite{Tichy03a,Tichy:2003qi,Ansorg:2004ds} we assume that $m$ 
is a good approximation for the initial individual black hole masses.

To complete the definition of the initial data, we also need to specify
initial values for the lapse $\alpha$ and shift
vector $\beta^i$. At time $t=0$ we use
\ba
\alpha &=& \left(1 + \frac{m_b}{4 r_1} + \frac{m_b}{4 r_2}\right)^{-4},\nonumber \\
\beta^i &=& 0, \nonumber
\ea
where $r_A$ is the coordinate distance from puncture $A$.
Lapse and shift evolve according to
\ba
(\partial_t - \beta^i\partial_i) \alpha &=& -2\alpha K , \nonumber \\
(\partial_t - \beta^k\partial_k)\beta^i &=& \frac{3}{4} B^i , \nonumber \\
(\partial_t - \beta^k\partial_k) B^i    &=&
  (\partial_t - \beta^k\partial_k)\tilde \Gamma^i - \eta B^i .\nonumber 
\ea

The gravitational fields are evolved using the ``moving punctures" method
\cite{Campanelli:2005dd,Baker:2005vv} with the implementation discussed
in~\cite{Bruegmann:2006at,Marronetti:2007ya},
with the exception that as in~\cite{Tichy:2007hk} the
dynamical variable $\phi$ has been replaced by the variable
\ba
W = e^{-2\phi} , \nonumber 
\ea
which obeys the evolution equation
\ba
\partial_t W = \frac{1}{3} W \left( \alpha K - \partial_i \beta^i \right) +\beta^i\partial_i W .\nonumber 
\ea
The new variable $W$ has two advantages. First, our simulations seem to 
converge better when we use $W$ instead of $\phi$ or $\chi = e^{-4\phi}$  
\cite{Campanelli:2005dd}. This may be related to the fact that $W$ grows 
linearly near the black hole centers after some time of evolving the system. 
Such linear behavior leads to more accurate finite differencing derivatives 
near the punctures. In addition, the determinant of the
3-metric $\det(\gamma_{ij})=W^{-6}$ remains always positive,
even if $W$ becomes slightly negative due to numerical error.
This property is not ensured if one evolves, for example, the
variable $\chi$. In the latter case one has to explicitly guard against
this to prevent code crashes.

\begin{table}
\centering
\begin{tabular}{c|c}
$Grid$	& $Structure$  \\
\hline
\hline
1  & $[5\times 48:5\times 54]~[h_{min}=M/56.9 :OB=238.5M]$\\
2  & $[5\times 56:5\times 63]~[h_{min}=M/66.4 :OB=235.3M]$\\
3  & $[5\times 64:5\times 72]~[h_{min}=M/75.9 :OB=246.4M]$\\
4  & $[5\times 72:5\times 81]~[h_{min}=M/85.3 :OB=243.0M]$\\
5  & $[5\times 80:5\times 90]~[h_{min}=M/94.8 :OB=240.3M]$\\
6  & $[5\times 80:5\times 90]~[h_{min}=M/85.3 :OB=267.0M]$\\
L  & $[3\times 69:6\times 149]~[h_{min}=M/80.0 :OB=256.0M]$\\
\end{tabular}
\caption{\label{table_grids}
Characteristics of our numerical grids. The values between brackets show the 
number of inner (moving) levels times the number of grid points per level per 
dimension, plus the number of outer (fixed) levels times the number of grid 
points. $h_{min}$ is the finest grid spacing and $OB$ gives the coordinate 
distance to the grid's outer boundary. The last entry corresponds to a grid 
used with the code LEAN \cite{Sperhake:2006cy}.}
\end{table}

The simulations presented here were performed using the BAM code, details of
which are described in \cite{Bruegmann:2003aw,Bruegmann:2006at}. 
BAM is based on three-dimensional Cartesian coordinates and can achieve high
spatial resolution near the punctures using a structure of nested refinement
levels. The outermost of these levels are fixed, while the innermost track
the movement of the punctures. For the runs presented here, we used 10
levels of refinement with the outer boundaries located about $240M$ away
from the grid center. Since the black holes are identical, we use quadrant
symmetry. Table \ref{table_grids} lists the characteristics of the different
grid layouts used in this paper. BAM characteristics make it ideal for quick
and relatively inexpensive runs \cite{Marronetti:2007ya}. Most of the
simulations presented here were done on dual processor workstations with
characteristics speeds of 0.9, 1.7 and 2.5 days per orbit when using grids
1, 2 and 3 respectively.

\begin{table}
\vspace {7mm}
\centering
\begin{tabular}{c|c|c|c}
$Run$ & $M_f/M$ & $J_f/M^2$ & $J_f/M_f^2$ \\
\hline
\hline
$S-0.90\_1$ & 0.976 & 0.348 & 0.365 \\ 
$S-0.90\_2$ & 0.970 & 0.359 & 0.382 \\ 
$S-0.90\_3$ & 0.970 & 0.358 & 0.380 \\
$S-0.90\_4$ & 0.970 & 0.358 & 0.380 \\ 
\hline
$S-0.75\_1$ & 0.970 & 0.405 & 0.431 \\ 
$S-0.75\_2$ & 0.968 & 0.408 & 0.436 \\
$S-0.75\_3$ & 0.968 & 0.409 & 0.437 \\ 
\hline
$S-0.50\_1$ & 0.963 & 0.483 & 0.520 \\ 
$S-0.50\_2$ & 0.963 & 0.486 & 0.524 \\
\hline
$S-0.25\_1$ & 0.958 & 0.551 & 0.600 \\ 
$S-0.25\_2$ & 0.958 & 0.555 & 0.606 \\
\hline 
$S+0.00\_1$ & 0.951 & 0.605 & 0.669 \\
$S+0.00\_2$ & 0.951 & 0.614 & 0.679 \\ 
$S+0.00\_3$ & 0.951 & 0.619 & 0.684 \\ 
$S+0.00\_4$ & 0.951 & 0.619 & 0.684 \\ 
\hline
$S+0.25\_1$ & 0.944 & 0.671 & 0.753 \\ 
$S+0.25\_2$ & 0.944 & 0.676 & 0.759 \\
\hline
$S+0.50\_1$ & 0.934 & 0.721 & 0.827 \\
$S+0.50\_2$ & 0.933 & 0.720 & 0.826 \\
\hline
$S+0.62\_1$ & 0.927 & 0.731 & 0.851 \\ 
$S+0.62\_2$ & 0.926 & 0.737 & 0.859 \\
$S+0.62\_3$ & 0.926 & 0.737 & 0.859 \\
\hline
$S+0.75\_1$ & 0.917 & 0.729 & 0.866 \\
$S+0.75\_2$ & 0.916 & 0.733 & 0.874 \\
$S+0.75\_3$ & 0.916 & 0.736 & 0.877 \\
$S+0.75\_4$ & 0.916 & 0.738 & 0.880 \\
\hline
$S+0.82\_1$ & 0.911 & 0.688 & 0.828 \\
$S+0.82\_2$ & 0.910 & 0.703 & 0.850 \\
$S+0.82\_3$ & 0.909 & 0.701 & 0.848 \\
\hline
$S+0.90\_1$ & 0.905 & 0.645 & 0.788 \\
$S+0.90\_2$ & 0.905 & 0.640 & 0.781 \\
$S+0.90\_3$ & 0.903 & 0.643 & 0.788 \\
$S+0.90\_4$ & 0.902 & 0.643 & 0.790 \\
$S+0.90\_5$ & 0.902 & 0.646 & 0.794 \\
$S+0.90\_6$ & 0.906 & 0.644 & 0.785 \\

\end{tabular}
\caption{\label{table_runs}
Binary black hole simulations performed for this paper. The runs are named 
$SXX\_Y$, where $XX$ indicates the value of $s/m^2$ and $Y$ is the numerical 
grid used for that run. Table \ref{table_grids} describes the different grids.}
\end{table}
Table \ref{table_runs} enumerates the binary simulations using the nomenclature 
$SXX\_Y$, where $XX$ indicates the value of $s/m^2$ and $Y$ indicates which one 
of the numerical grids of Table \ref{table_grids} was used. We measure the mass 
and angular momentum of the remnant of the binary black hole merger using 
techniques based on volume integrals (see \cite{Marronetti:2007ya} for details). 
These values of mass and angular momentum are labeled $M_f$ and $J_f$ 
respectively. 

\begin{figure}
\includegraphics[scale=0.34,clip=true]{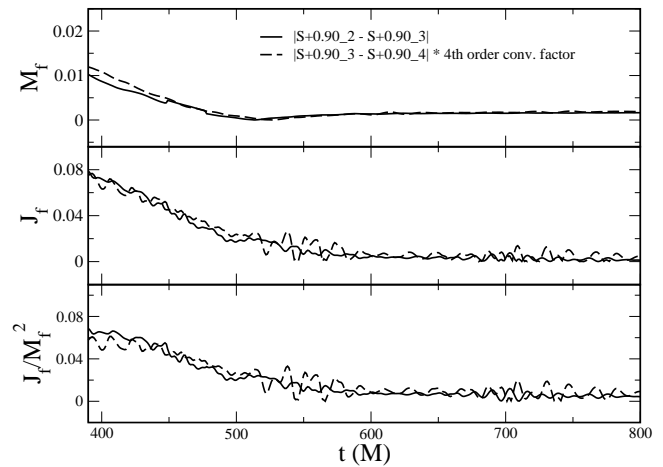}
\caption{\label{Fig_Conv}
Mass and angular momentum plots for three different resolutions. The dashed 
curves have been scaled according to a factor corresponding to 4th order 
convergence in spatial resolution.}
\end{figure}

We study the effect of different approximations and limitations inherent to our 
simulations: grid resolution, grid structure, measurement of the final mass and 
angular momentum, and the intrinsic characteristics of our initial data sets. We 
start by varying the spatial resolution of our grids, while keeping the grid 
size and structure (i.e., the layout of the refinement levels) unchanged. Grids 
1 to 5 cover maximum spatial resolutions going from $M/56.9$ to $M/94.8$. 
The results, presented in Table \ref{table_runs}, do not seem to vary 
significantly with the spatial resolution: the values of $J_f/M_f^2$ obtained 
for grids 2 to 5 agree with each other at a level of about $1.5\%$ or better. 
We test the convergence of the calculations of mass and angular momentum, using 
four runs for the binary with spin $s/m^2=0.90$ (labeled $S+0.90\_Y$, $Y$=1 to 
4). These runs shared the same grid layout but with varying spatial resolution, 
ranging from $h_{min}= M/56.9$ to $h_{min}= M/85.3$ \footnote{Run S+0.90\_5 was 
not used because the high order of our algorithms (fourth) made the difference 
in results fall too close to our calculation round-off error for meaningful 
convergence tests.}. We find that the lowest resolution run ($S+0.90\_1$) fell 
outside the convergence regime. The results of the other three runs, presented 
in Fig. \ref{Fig_Conv}, seem to show that these runs are in the convergent 
regime. 

To evaluate the influence of the grid layout, we compare the results from two 
grids with identical resolution but different number of grid points per box 
(Grids 4 and 6 from Table \ref{table_grids}). The box size has been found 
crucial in previous work \cite{Bruegmann:2006at}: boxes smaller than critical 
sizes tend to change drastically the binary dynamics, altering orbits and 
merger times. Runs $S+0.90\_4$ and $S+0.90\_6$ returned values of mass and 
angular momentum with differences of $\Delta M_f=4~10^{-3} (0.4\%)$ and $\Delta 
J_f=5~10^{-4} (0.1\%)$ respectively. 

Additionally, and as an independent check, we compare our results with those of 
Campanelli \et ~\cite{Campanelli:2006uy} (shown in Fig. \ref{Fig_JM2} as empty 
circles) and Pollney \et ~\cite{Pollney:2007ss} (empty square).

To evaluate the accuracy of the algorithms used to measure mass and angular 
momentum, we compare our volume-integral based results with calculations done 
using surface integrals \cite{Marronetti:2007ya} and find differences of up 
to $0.5\%$ in magnitude. We also test the satisfaction of the Christodoulou 
formula $J_c=2~M_{irr}~(M_f^2-M^2_{irr})^{1/2}$ ($M_{irr}$ being the irreducible 
mass of the final black hole). The relative difference in the values of $J_f$ 
and $J_c$ for the run $S+0.90\_5$ was less than $0.5\%$.

\section{Results}
\label{results}

\begin{figure}
\includegraphics[scale=0.34,clip=true]{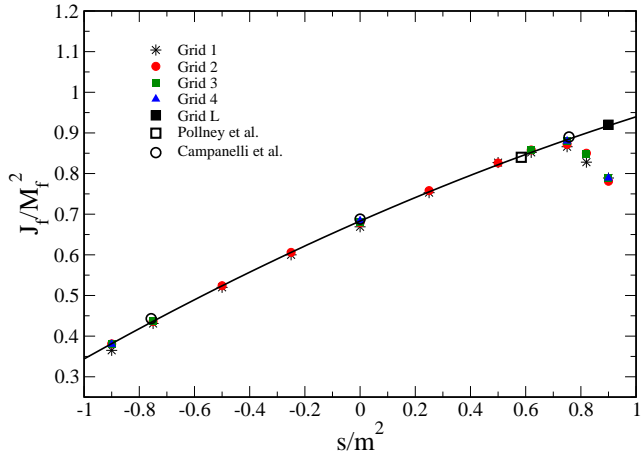}
\caption{\label{Fig_JM2}
Spin of the merger remnant as a function of the initial black hole spins 
measured $500M$ after the black hole merger. The binaries are composed of 
identical black holes with initial spins of magnitude $s/m^2$ aligned with the 
orbital angular. The values calculated in \cite{Campanelli:2006uy} 
(\cite{Pollney:2007ss}) are shown as empty circles (square). The solid square 
represents the remnant spin calculated right after the merger for the case 
$s/m^2=0.90$, using Grid L. The solid line is a quadratic interpolation of the 
values corresponding to Grid 2 for $s/m^2<0.75$ plus the Grid L value.}
\end{figure}

\begin{figure}
\includegraphics[scale=0.34,clip=true]{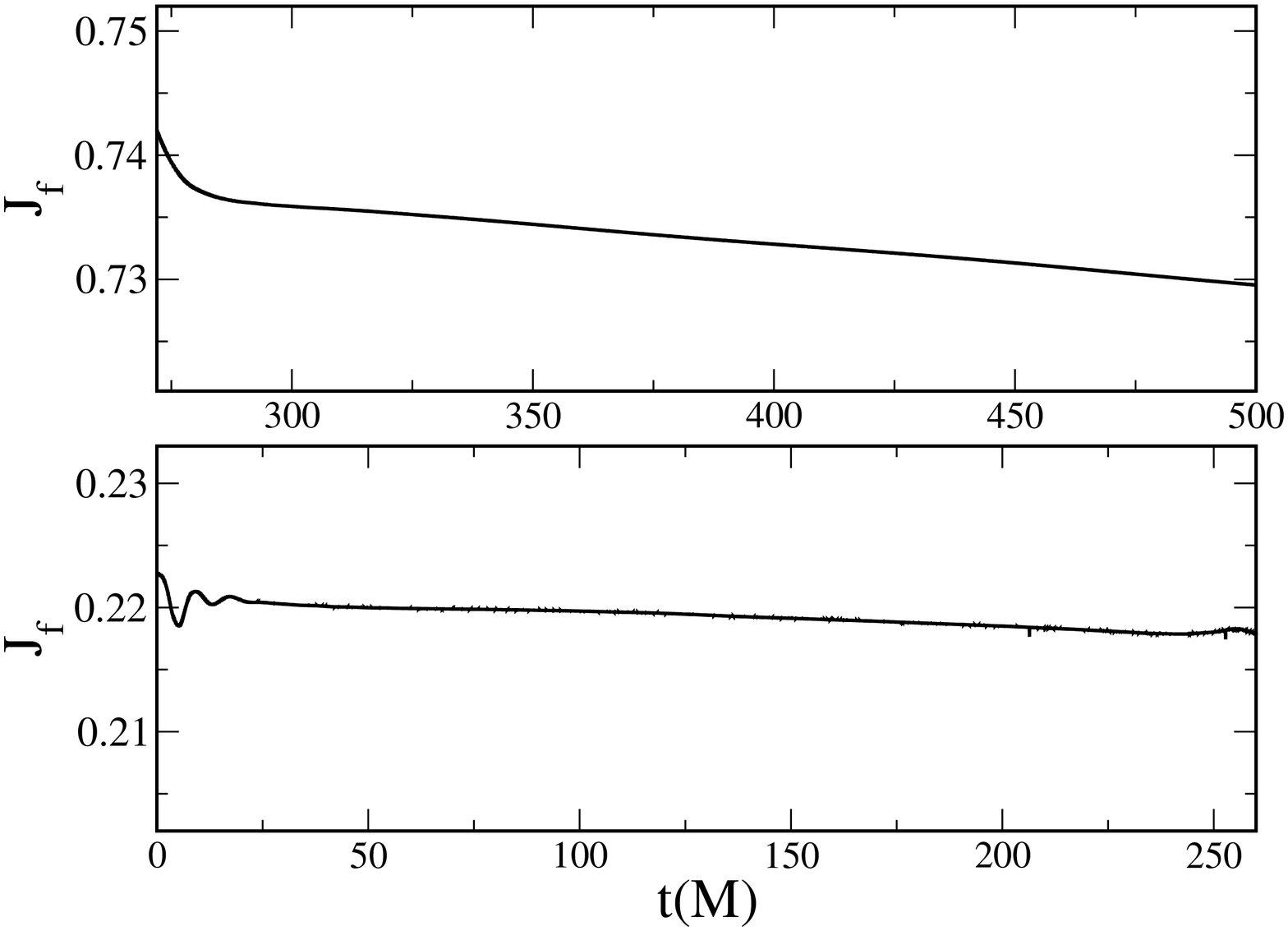}
\caption{\label{Fig_S0.90_iso}
Evolution of the angular momentum for the binary with $s/m^2=0.90$ obtained 
using the isolated horizon techniques of \cite{Campanelli:2006fy}. The top plot 
corresponds to the angular momentum of the remnant of the merger, while the 
bottom plot tracks the spin of the individual black holes before the merger 
(occurring at $265M$).}
\end{figure}

Figure \ref{Fig_JM2} summarizes the results of Table \ref{table_runs}. The
spin parameter of the binary remnant $J_f/M_f^2$ is shown as a function of
the initial black hole spins $s/m^2$. These values were measured at $500M$
after the merger (which corresponds approximately to the side-to-side
light-crossing time of our grids).
The most striking feature of the plot is the existence of an apparent
maximum at $s/m^2 \approx 0.75$. 
As we will see, this maximum is merely a numerical artifact caused
by an artificial loss of angular momentum in the 
case of highly spinning black holes if we wait for $500M$ after the merger.
Shortly after the merger, the spin of the remaining black hole is
still larger. This is confirmed by a simulation of the $s/m^2=0.90$
case using the LEAN code (Grid $L$ in Table \ref{table_grids}) that tracks
the common apparent horizon and the emission of energy and angular momentum
in the form of gravitational waves. We calculate the spin about $50M$ after
the merger in four different ways: from the balance of
gravitational wave energy and angular momentum loss, from the ringdown
frequencies (using the tabulated results given in \cite{Berti:2005ys}), from
the isolated horizon geometry (using the techniques from
\cite{Campanelli:2006fy}), and using Christodoulou's formula. The
corresponding results are $J_f/M^2_f=0.95,~0.925,~0.913,~0.918$ respectively. 
The first result is the least accurate, given that the wave extraction radius 
($50M$) was not considerably
far from the center. The rest cluster around 0.92 (plotted in Fig. 
\ref{Fig_JM2} as a solid square) that agrees better with an extrapolation from
lower $s/m^2$ results. Figure \ref{Fig_S0.90_iso} shows the evolution of
the angular momentum obtained using isolated horizon techniques which
indicates a loss rate of about $0.5\%$ per $100M$. This last result
together with the measurements at $500M$ after the merger seem to indicate 
an artificial loss in angular momentum for high spin black holes. 
Also note that this effect is not seen at the
lower spin end of Fig. \ref{Fig_JM2}. This is simply due to the fact that the
merger remnant of binaries with individual spins $s/m^2 < 0.75$ is a low
spinning black hole, and thus not affected by this effect.

\begin{figure}
\includegraphics[scale=0.34,clip=true]{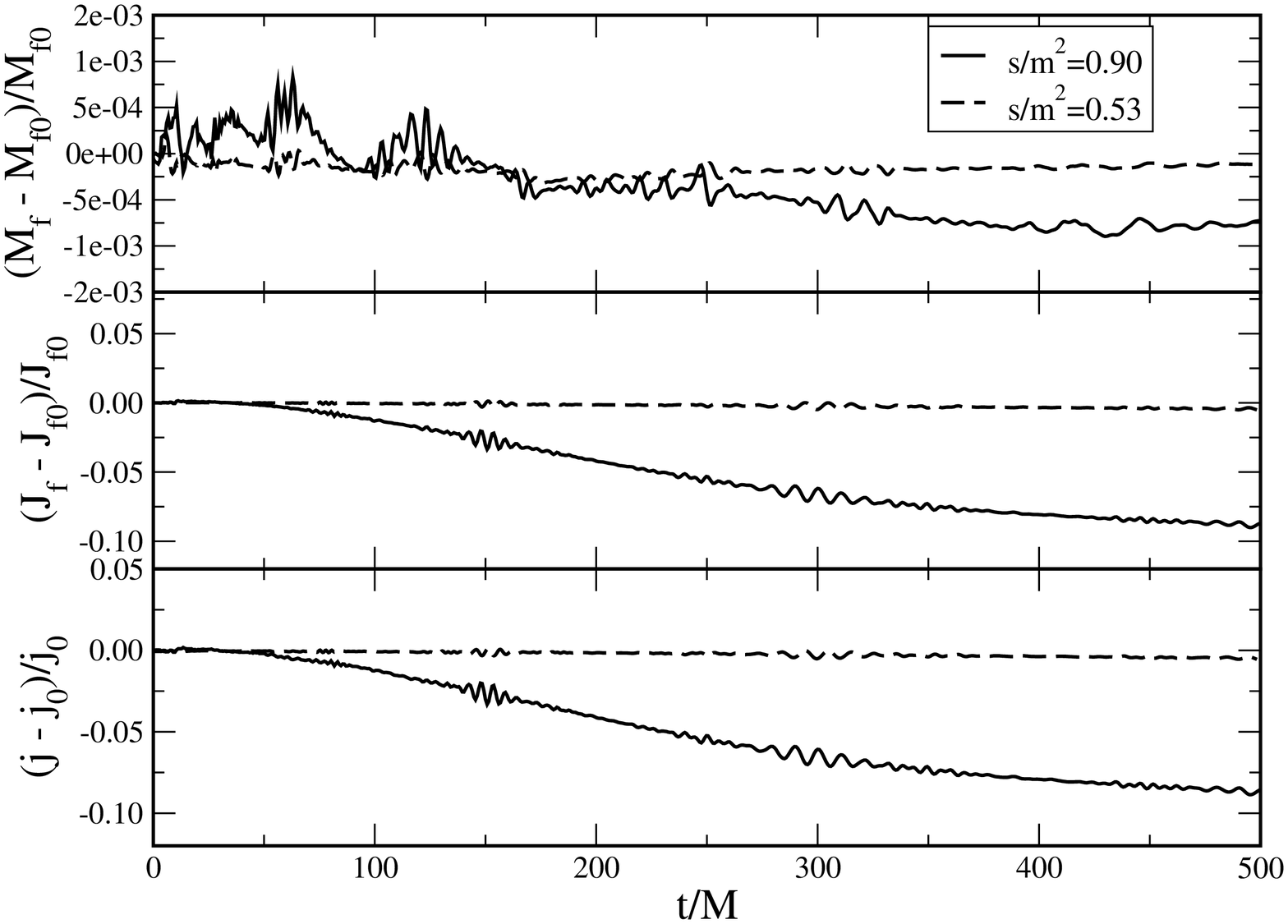}
\caption{\label{Fig_SBH_Comp_Diff}
Single black hole evolutions with $s/m^2=0.53$ (dashed) and $0.90$ (solid), on 
Grid 4. The plots present the relative differences with respect to the initial 
values, denoted with the $0$ sub-index. The bottom plot presents the relative 
variation of intrinsic angular momentum $j\equiv J_f/M^2_f$.}
\end{figure}

\begin{figure}
\includegraphics[scale=0.34,clip=true]{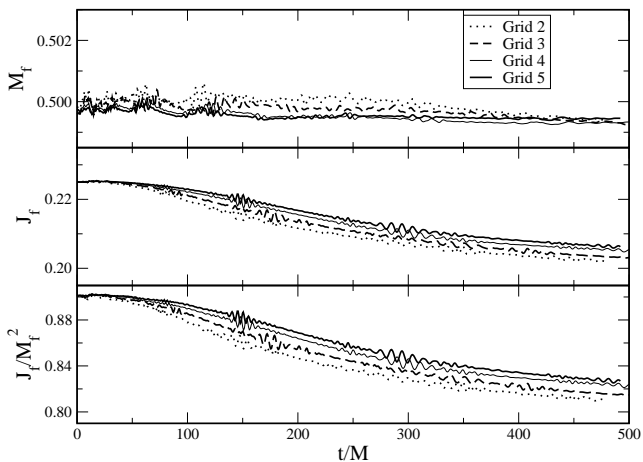}
\caption{\label{Fig_SBH}
Evolution of single black holes with $s/m^2=0.90$ for different 
spatial resolutions.}
\end{figure}

To study this effect in more detail, we perform single black hole evolutions. 
Figure \ref{Fig_SBH_Comp_Diff} compares evolutions for the cases with 
$s/m^2=0.53$ and $0.90$. In order to facilitate the comparison, we plot the 
relative differences from the initial values of the mass, angular momentum and 
intrinsic angular momentum parameter, these
quantities measured using volume integrals methods. The angular momentum loss, 
while negligible in the $0.53$ case, is more pronounced in the larger spin case. 
Note however that the mass is largely unaffected in both cases.
Figure \ref{Fig_SBH} shows the results for four different grid resolutions,
indicating firstly that the mass is better conserved than the angular
momentum and secondly that this loss gets smaller with increasing
resolution, albeit quite slowly. 
The angular momentum curves are still consistent with 4th order convergence, 
however, Richardson extrapolating these curves to arbitraily large resolution 
still shows loss of angular momentum.
For BBH evolutions with many orbits before
the merger, this loss could in principle produce an ``effective" value of
$s/m^2$ at the time of the merger smaller than the initially intended. In
order to evaluate this effect, we track the evolution of the individual
spins before the merger as shown in the bottom panel of Fig.
\ref{Fig_S0.90_iso}. An angular momentum loss of about $3\%$ is detected
before the merger, which occurs at $t=265M$.

At first sight, part of this loss could also be attributed to our choice of
initial data. Traditionally, puncture initial data sets are constructed by
solving the Hamiltonian constraint under the assumption of conformal
flatness for the spatial metric. The momentum constraint is analytically
satisfied by the use of the Bowen-York (BY) formula for the extrinsic
curvature which, in principle, allows for the arbitrary specification of the
linear and angular momentum of the black holes. In practice, the amount of
``junk radiation" associated with these initial data sets increases with the
magnitude of the spin of the hole. The amount of this extra radiation has
been studied perturbatively by Gleiser \et ~\cite{Gleiser:1997ng} and
numerically by Hannam \et ~\cite{Hannam:2006zt} and 
Dain \et ~\cite{Dain:2002ee}. 
The last shows that, for the case $s/m^2=0.90$, a
single BY black hole would relax into a Kerr black hole after
emitting about $0.1\%$ of the initial gravitational mass.

\begin{figure}
\includegraphics[scale=0.34,clip=true]{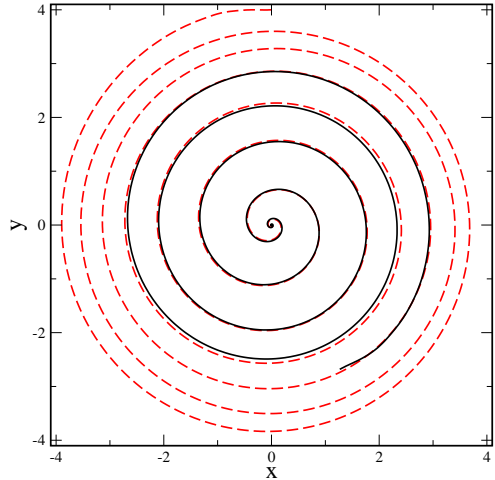}
\caption{\label{Fig_Large_Sep_orbits}
Puncture tracks of one of the holes for two simulations with $s/m^2=0.90$ that 
start out at different coordinate separations; $D=8M$ (dashed) and $D=6M$ 
(solid). The $D=6M$ orbits have been rotated to coincide with the ones from the 
$D=8M$ case in the final orbits. Both runs were performed using Grid 3.}
\end{figure}

\begin{figure}
\includegraphics[scale=0.34,clip=true]{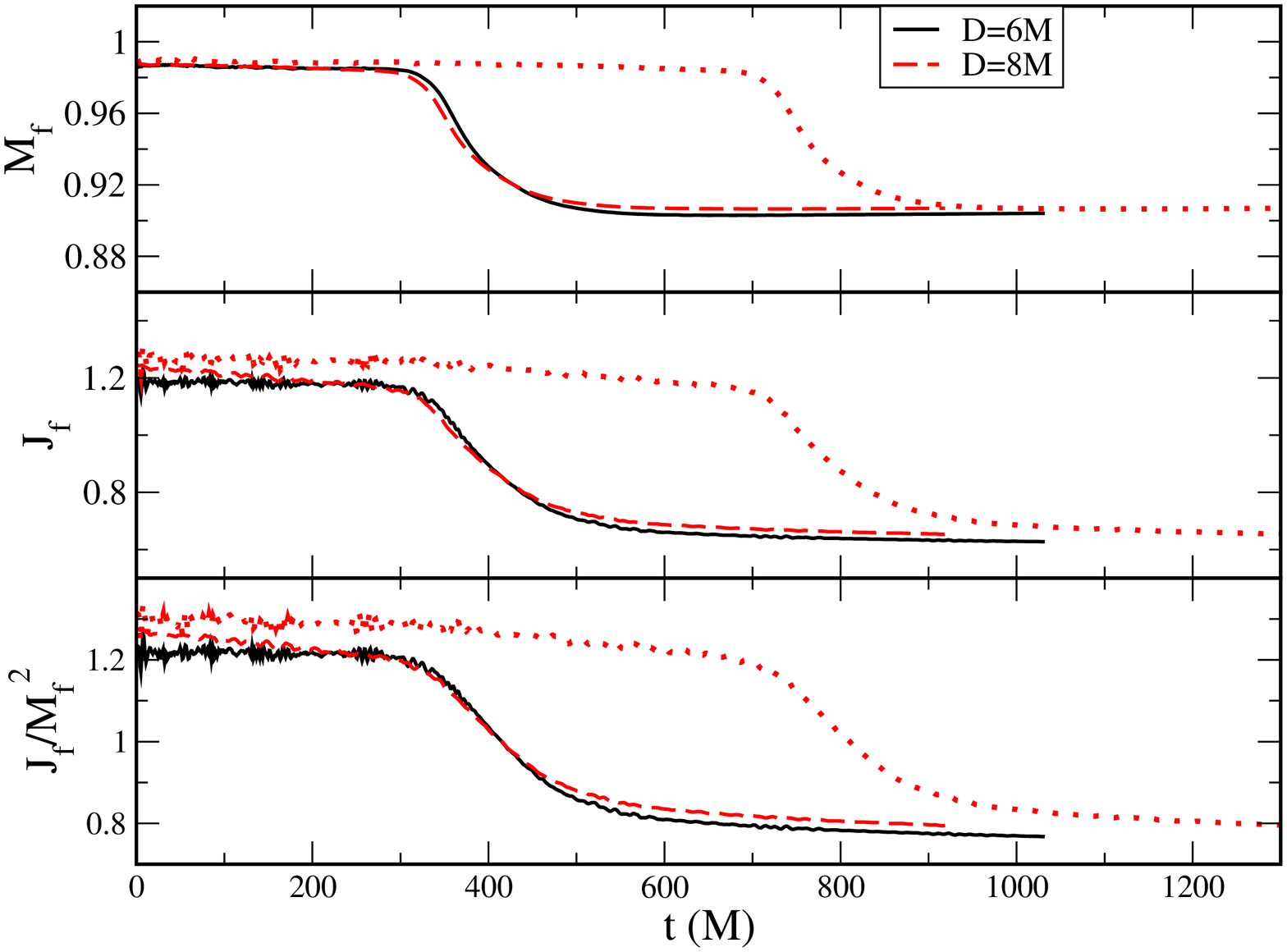}
\caption{\label{Fig_Large_Sep}
Evolution of mass and angular momentum for the runs of Fig. 
\ref{Fig_Large_Sep_orbits}. The solid (dotted) lines corresponds to the 
separation $D=6M$ ($D=8M$). The dashed line corresponds to the $D=8M$ run 
shifted in time by $395M$.}
\end{figure}

To further study  this effect in the evolution prior to the merger, we
perform a simulation of a binary with $s/m^2=0.90$ that starts out at a
coordinate separation $D=8M$ and compare it with one starting at $D=6M$. The
former simulation lasted more than two orbits and a half longer than the latter,
corresponding to an additional simulation time of $395M$. Figure
\ref{Fig_Large_Sep_orbits} shows the orbits of one of the punctures for
these simulations. Both runs were performed using Grid 3. Figure
\ref{Fig_Large_Sep} shows the corresponding evolution of the mass and
angular momentum, where the curves corresponding to the $D=8M$ run were
shifted in time by $395M$. Naively, one would expect the $D=8M$ to dissipate
more angular momentum before the merger, given that it is more than two
orbits longer than its $D=6M$ counterpart. This would result in the longer
run remnant having a smaller intrinsic spin than the shorter evolution.
However, the resulting spin for the $D=8M$ run is about $3\%$ larger for the
$D=6M$ case, but within the error margin for these measurements, making it
difficult to draw any conclusions.

\begin{table}
\vspace {7mm}
\centering
\begin{tabular}{l|c|c|c|c}
~~ & {\bf C} & {\bf R} & {\bf M} & {\bf R+M} \\
\hline
\hline
$p_0$ 	& 0.6879 & 0.6883(4)  & 0.6855(16) & 0.6888(4) \\ 
$p_1$ 	& 0.1476 & 0.1530(4)  & 0.1499(8)  & 0.1525(5) \\ 
$p_2$	&-0.0093 &-0.0088(5)  &-0.0110(8)  &-0.0106(5) \\
Max $J/M^2$  & 0.946 & 0.959(2) & 0.941(6) & 0.951(4) \\
Min $J/M^2$  & 0.355 & 0.347(2) & 0.342(6) & 0.341(4)
\end{tabular}
\caption{\label{table_fit_Rezz}
Least-squares fit of Eq. (\ref{Fit_Rezzolla}) from Campanelli \et 
~\cite{Campanelli:2006fy} ({\bf C}), Rezzolla \et ~\cite{Rezzolla:2007xa} 
({\bf R}), this paper ({\bf M}), and Rezzolla \et ~and our data sets 
combined ({\bf R+M}). The last two rows show the extrapolation to initial 
critical black holes aligned (counter-aligned) with the orbital angular 
momentum that corresponds to the maximum (minimum) possible intrinsic 
spin for the remnant black hole.}
\end{table}

Next, we fit the values of the highest resolution runs from Table 
\ref{table_runs}, ignoring the results for $s/m^2 > 0.75$, but adding the 
result at $s/m^2 =0.90$ 
from Grid L (solid square). We follow the fitting formula for the final 
black hole intrinsic spin used in Campanelli \et ~\cite{Campanelli:2006fy}
and Rezzolla \et~ \cite{Rezzolla:2007xa} (Eq. (8))
\ba
J_f/M_f^2 = p_0 &+& p_1~(s_1/m_1^2+s_2/m_2^2) \nonumber \\
	&+& p_2~(s_1/m_1^2+s_2/m_2^2)^2,
\label{Fit_Rezzolla}
\ea
where $p_0, ~p_1$ and $p_2$ are the fitting parameters. We present in Table 
\ref{table_fit_Rezz} a comparison of our best fit parameters with those from 
\cite{Campanelli:2006fy} and \cite{Rezzolla:2007xa} and with the ones obtained 
by fitting \cite{Rezzolla:2007xa} and our data sets together.
For these fits, we assume a nominal error of $0.02$ for all our values. We see 
that the fitting parameters are in an agreement consistent with small-number
statistics in all cases. Table \ref{table_fit_Rezz} also shows the extrapolation 
to maximally rotating black holes, aligned and counter-aligned with the orbital 
angular momentum. The maximum and minimum values for the intrinsic 
angular momentum of the remnant predicted by the fit of the 
combined data sets is $J_f/M_f^2=0.341 \pm 0.004$ and $0.951 \pm 0.004$ 
respectively \footnote{To asses the influence of the LEAN data point in our 
results, we re-calculated our predictions for the maximum and minimum values
of $J_f/M_f^2$ without it. The new results are within the error bands of 
the values quoted in Table \ref{table_fit_Rezz}}. Note that, while our data 
set is smaller than the one in \cite{Rezzolla:2007xa}, it
contains measurements that are closer to the extrapolated values for critical 
BBH.

Recently, Boyle \et~ \cite{Boyle:2007sz} 
introduced a formalism that predicts any final quantity resulting from the merger
using a Taylor expansion on the initial binary mass ratio $q \equiv m_1/m_2$ and
the components of the initial spins. For the highly symmetric binaries 
considered in this paper, their expansion for $J_f/M_f^2$ to second order 
reduces to the polynomial of Eq. (\ref{Fit_Rezzolla}) with the equivalences 
$p_0=s_3^{000|000}$, $p_1=s_3^{001|000}$ and 
$p_2 = (1/4)~ (2~ s_3^{002|000} + s_3^{001|001})$, where the 
parameters on the right-hand sides are those from the corresponding expansion 
in \cite{Boyle:2007sz}.

Finally, we would like to highlight the ability of the code BAM to probe 
accurately and efficiently BBH parameter space. Our lowest 
resolution runs (Grid 1), performed on dual-processor workstations, are good 
enough to capture the main characteristics of the results of Fig. \ref{Fig_JM2}.

\section{Conclusions}
\label{conclusions}

We studied the effect of the initial spins of the black holes in a binary 
system on the mass and angular momentum of the black hole that results from 
the merger. We concentrated on equal-mass binaries with spins aligned with the 
orbital angular momentum ($s_1 = s_2$), covering a range of initial spin 
parameters going from $s/m^2=-0.90$ to $0.90$. The runs at the extrema of the 
range are the highest spin simulations to date. The main results of the paper 
are presented in Fig. \ref{Fig_JM2}, where the spin parameter of the remnant 
($J_f/M_f^2$) is given as a function of the initial spin parameters. 

We combined our results with those of Rezzolla \et 
~\cite{Rezzolla:2007xa} in a quadratic least-square fit and obtain, 
by extrapolation to initial critical black holes, predicted 
maximum and minimum values of $J/M^2$ for the black hole 
remnant of $0.951 \pm 0.004$ and $0.341 \pm 0.004$ respectively. 
These error bounds are simply derived from the uncertainty of the
fitting parameters provided by their standard error. The small
size of the samples studied here plus the fact that the limits to
critical black hole results are obtained through extrapolation
to points in parameter space with singular properties may
lead to revisions in these estimates once binaries 
with spins larger than 0.90 can be accurately simulated.
For this, new recipes for initial data sets that allow 
for initial black holes with spin parameters larger than 0.928 
(the limit of BY data) are needed. The methods introduced by Dain 
\cite{Dain00,Dain01b} and studied in head-on BBH simulations by 
Hannam \et ~\cite{Hannam:2006zt} may hold the key to this problem.

We also find  a problem for the simulations starting with spins $s/m^2 > 0.75$. 
Current evolutions based on the moving punctures methods present losses of 
angular momentum at non-negligible rates for highly 
spinning black holes, even when relatively high grid resolutions are employed. 
Measurements of the merger remnant intrinsic spin $500M$ after the merger of 
two $s/m^2=0.90$ black holes show the spurious loss of more than $10\%$ of the 
angular momentum. This effect increases with the magnitude of the black hole 
spin. This loss does not affect strongly the calculation of gravitational wave 
templates, since they cover only up to a short period after the merger. However, 
it becomes more important, for instance, for simulations of black hole-neutron 
star or neutron star-neutron star binaries, where the potential formation of an 
accretion disk around a black hole has to be followed for much longer periods.

\begin{acknowledgments}

It is a pleasure to thank to Mark Hannam and Sascha Husa for their help and 
insight all along this project. We would also like to thank C. Lousto, 
L. Boyle and M. Kesden for useful discussions. 
This work was supported by NSF grants PHY-0555644 and PHY-0652874. 
We also acknowledge
partial support by the National Computational Science Alliance under
Grants PHY050016N and PHY060021P and by DFG grant SFB/Transregio~7 
``Gravitational Wave Astronomy''. We thank the DEISA Consortium 
(co-funded by the EU, FP6 project 508830), for support within the 
DEISA Extreme Computing Initiative (www.deisa.org). J.G. and U.S. acknowledge 
support from the ILIAS Sixth Framework Programme. The high resolution 
simulations were performed at the Charles E. Schmidt College of Science 
computer cluster {\it Boca 5}, at the Cray XT3 MPP system (BigBen) at the 
Pittsburgh Supercomputer Center, and at LRZ Munich. \\

\end{acknowledgments}

\bibliography{references}

\end{document}